\documentclass[journal=jpccck,manuscript=article,layout=twocolumn]{achemso}

\setkeys{acs}{articletitle=true}
\usepackage{lettrine}

\usepackage[version=3]{mhchem} 
\usepackage[T1]{fontenc}       

\usepackage[fontsize=10pt]{scrextend}

\usepackage{newtxtext}
\usepackage{bm} 
\usepackage[labelfont=bf]{caption} 

\usepackage{titlesec} 
\titleformat{\subsection}[runin]
{\normalfont\bfseries}{\thesubsection{.}}{1em}{}[.]

\titleformat{\section}
{\normalfont\sffamily\large\bfseries}{\thesection{.}}{1em}{\MakeUppercase}

\usepackage{xcolor}
\usepackage{graphicx}
\usepackage{amsmath,fullpage,amssymb}
\usepackage{graphics,epsfig}
\usepackage{gensymb}

\def\Eq{eq}

\def\Fig{Figure}
\def\Figs{Figures}

\usepackage{soul}	

\usepackage[load=abbr]{siunitx}
\usepackage{subfig}

\usepackage{xspace}
\usepackage{xr}
\def\Fig{Fig.}
\def\Figs{Figs.}
\def\Eq{Eq.}

\newcommand*{\citen}[1]{
  \begingroup
    \romannumeral-`\x 
    \setcitestyle{numbers}%
    \cite{#1}%
  \endgroup   
}

\newcommand{\GDME}{${g}_{\text{\ce{Li+}--DME}}(r)$\xspace}

\newcommand{\GLiTFSI}{${g}_{\text{\ce{Li+}--\ce{TFSI-}}}(r)$\xspace}

\newcommand{\Glisx}{${g}_{\text{\ce{Li+}--\ce{S_$x$^{2-}}}}(r)$\xspace}

\newcommand{\sx}{\ce{S_{$x$}^{2-}}\xspace}

\newcommand{\liion}{\ce{Li^{+}}\xspace}
\newcommand{\tfsiion}{\ce{TFSI^{-}}\xspace}

\newcommand{\seight}{\ce{S8^{2-}}\xspace}
\newcommand{\ssix}{\ce{S6^{2-}}\xspace}

\newcommand{\sfour}{\ce{S4^{2-}}\xspace}

\newcommand{\lseight}{\ce{Li$_2$S$_8$}\xspace}
\newcommand{\lssix}{\ce{Li$_2$S$_6$}\xspace}
\newcommand{\lsfour}{\ce{Li$_2$S$_4$}\xspace}
\newcommand{\lstwo}{\ce{Li$_2$S$_2$}\xspace}
\newcommand{\lsone}{\ce{Li$_2$S}\xspace}

\newcommand{\lsx}{\ce{Li$_2$S$_x$}\xspace}

\newcommand{\etal}{$et$ $al.$\xspace}

\newcommand{\mrm}[1]{{\mathrm{#1}}}
\newcommand{\trm}[1]{{\textrm{#1}}}

\SectionNumbersOn
\let\oldmaketitle\maketitle
\let\maketitle\relax

\title{\flushleft 
Structural and Transport Properties of Li/S Battery Electrolytes: Role of the Polysulfide Species}

\author{Chanbum Park}
\affiliation[hzb]{\rm\small Research Group for Simulations of Energy Materials, Helmholtz-Zentrum Berlin f\"ur Materialien und Energie, Hahn-Meitner-Platz 1, D-14109 Berlin, Germany}
\alsoaffiliation[hu]
{\rm\small Institut f\"ur Physik, Humboldt-Universit\"at zu Berlin, Newtonstr.~15, D-12489 Berlin, Germany}

\author{Arne Ronneburg}
\affiliation{\rm\small Institut f\"ur Weiche Materie und Funktionale Materialien, Helmholtz-Zentrum Berlin f\"ur Materialien und Energie, Hahn-Meitner-Platz 1, D-14109 Berlin, Germany}
\alsoaffiliation
{\rm\small Institut f\"ur Physik, Humboldt-Universit\"at zu Berlin, Newtonstr.~15, D-12489 Berlin, Germany}

\author{Sebastian Risse}
\affiliation{\rm\small Institut f\"ur Weiche Materie und Funktionale Materialien, Helmholtz-Zentrum Berlin f\"ur Materialien und Energie, Hahn-Meitner-Platz 1, D-14109 Berlin, Germany}

\author{Matthias Ballauff}
\affiliation
{\rm\small Institut f\"ur Weiche Materie und Funktionale Materialien, Helmholtz-Zentrum Berlin f\"ur Materialien und Energie, Hahn-Meitner-Platz 1, D-14109 Berlin, Germany}
\alsoaffiliation
{\rm\small Institut f\"ur Physik, Humboldt-Universit\"at zu Berlin, Newtonstr.~15, D-12489 Berlin, Germany}

\author{Matej Kandu\v{c}}
\affiliation{\rm\small Research Group for Simulations of Energy Materials, Helmholtz-Zentrum Berlin f\"ur Materialien und Energie, Hahn-Meitner-Platz 1, D-14109 Berlin, Germany}
\alsoaffiliation{\rm\small Jo\v{z}ef Stefan Institute, Jamova 39, SI-1000 Ljubljana, Slovenia}

\author{Joachim Dzubiella}
\affiliation{\rm\small Applied Theoretical Physics -- Computational Physics, Physikalisches Institut, Albert-Ludwigs-Universit\"at Freiburg, Hermann-Herder Strasse 3, D-79104 Freiburg, Germany}
\alsoaffiliation
{\rm\small Research Group for Simulations of Energy Materials, Helmholtz-Zentrum Berlin f\"ur Materialien und Energie, Hahn-Meitner-Platz 1, D-14109 Berlin, Germany}
\email{joachim.dzubiella@physik.uni-freiburg.de}

\begin{document}
\pagenumbering{arabic}
\noindent

\parindent=0cm
\setlength\arraycolsep{2pt}

\twocolumn[	
\begin{@twocolumnfalse}
\oldmaketitle

\begin{abstract}
Lithium--sulfur (Li/S) batteries are regarded as one of the most promising energy storage devices beyond lithium-ion batteries because of their high energy density of \SI{2600}{\watt\hour\per\kilo\gram} and an affordable cost of sulfur. Meanwhile, some challenges inherent to Li/S batteries remain to be tackled, for instance, the polysulfide (PS) shuttle effect, the irreversible solidification of Li$_2$S, and the volume expansion of the cathode material during discharge. 
On the molecular level, these issues originate from the structural and solubility behavior of the PS species in bulk and in the electrode confinement.  In this study, we use classical molecular dynamics (MD) simulations to  develop a working model for PS of different chain lengths in applied electrolyte solutions of lithium bistriflimide (LiTFSI) in 1,2-dimethoxyethane (DME) and 1,3-dioxolane (DOL) mixtures. We investigate conductivities, diffusion coefficients, solvation structures, and clustering behavior and verify our simulation model with experimental measurements available in literature and newly performed by us. Our results show that diffusion coefficients and conductivities are significantly influenced by the chain length of PS. The conductivity contribution of the short chains, like \sfour, is lower than of longer PS chains,
such as \ssix or \seight, despite the fact that the diffusion coefficient of \sfour is higher than for longer PS chains. The low conductivity of \lsfour can be attributed to its low degree of dissociation and even to a formation of large clusters in the solution. It is also found that an addition of 1 M LiTFSI into PS solutions considerably reduces the clustering behavior. Our simulation model enables future systematic studies in various solvating and confining systems for the rational design of Li/S electrolytes.
\\\\
\vspace{5ex}
\end{abstract}	
\end{@twocolumnfalse}]

\maketitle
\setlength\arraycolsep{2pt}
\section{Introduction}
Lithium-ion batteries are the dominant type of rechargeable batteries currently on the market. Relatively efficient battery life cycle, decent energy and power densities make lithium-ion batteries attractive. Meanwhile, the growing demand for large-scale energy storage systems and electric vehicles motivates the battery community to seek new solutions with higher energy density and lower material costs. One of the candidates are lithium--sulfur (Li/S) batteries owing to a low cost of sulfur, high energy density of \SI{2600}{\watt\hour\per\kilo\gram}, and high theoretical specific capacity of \SI{1675}{\ampere\hour\per\kilo\gram}.~\cite{fang2017more,kang2016review}

In spite of these attractive advantages, a practical Li/S battery performance has not been realized.  Unlike lithium-ion batteries, where intercalation of \liion provides the major dynamics, elemental sulfur S$_8$ in the cathode chemically reacts with \liion.~\cite{schipper2016brief, wild2015lithium} During discharge, \liion\! migrates from the lithium anode through the electrolyte and the separator to the sulfurized cathode. Subsequently, solid sulfur converts into linear chain-like polysulfide (PS) intermediates, whose length disproportionately shrinks upon reduction with lithium until Li$_2$S finally precipitates.~\cite{kang2016review, manthiram2014rechargeable, wild2015lithium, cuisinier2013sulfur} 
During charging, the reverse chemical reaction partially takes place. As a result of a complex cell chemistry, Li/S cells have to face two major challenges. First, parasitic side reactions at the lithium electrode can occur that originate from irreversible reduction of PS to lithium sulfide.~\cite{dominko2015analytical, wild2015lithium, cuisinier2013sulfur} This {\sl shuttle} phenomenon, originating from the migration of PS from the cathode materials toward the anode, is one source of capacity fading. Second, during discharge \lstwo increase in volume by 80\%, which can lead to mechanical stress in the cathode material and loss of contact to the conductive host matrix~\cite{fang2017more,busche2014systematical,mikhaylik2004polysulfide}. These adverse phenomena are the major bottlenecks of the Li/S battery development. As indicated by recent work,~\cite{cleaver2018perspective, kang2016review} extensive efforts have been devoted into minimizing the shuttling of the PS by confining them into meso- or micro-porous materials.~\cite{ji2010advances, wang2012microporous,zhang2010enhancement,su2012lithium, li2018revisiting} Because the soluble intermediate PS species cause the shuttling, the reaction mechanism pathway can also be altered by choices of the solvent and ions and thus changing the stability of the intermediate species, the charge/discharge rate, etc.~\cite{wild2015lithium, lee2017directing, scheers2014review, chen2018superior, li2018revisiting, urbonaite2014importance, fan2016solvent} Therefore, more fundamental and systematical studies on the structural and dissolution properties of PS in various applied solvents are in urgent need.

The most frequently formed intermediates during the battery cycling process are \sx (with $x=4,\ldots,8$). 
These PS ionic chains in the electrolyte can merge together with \liion\! into clusters, such as (Li$_2$S$_x$)$_n$ (with $n \geq 1$).~\cite{wang2014formation}  The clusters of Li$_2$S$_4$ are believed to be the last polysulfide intermediates before the formation of Li$_2$S$_2$/Li$_2$S insoluble aggregates during the discharge process.~\cite{partovi2015evidence} 
In recent years, various quantum chemistry calculations and classical MD models 
demonstrated the poor solubility and thus the tendency for cluster formation of
shorter PS species~\cite{wang2014formation, pascal2017thermodynamic, vijayakumar2014molecular, yu2017understanding, partovi2015evidence, kamphaus2017first, liu2018revealing,rajput2017elucidating, osellamodelling}.

The solubility of PS, the size distribution of the PS clusters and their morphologies depend on the properties of the solvent and added electrolyte,\cite{park2013solvent, zhang2014chelate, pan2015way, zheng2015quantitative} current density, and temperature.~\cite{yu2018direct} 
 A recent study by Andersen \etal~\cite{andersen2019structure} investigated PS clustering using classical MD simulations, DFT calculations and experiments. Another classical MD simulations by Rajput \etal~\cite{rajput2017elucidating} suggested that introducing \tfsiion ions, which compete with PS for \liion, weakens the PS -- \liion clustering networks, resulting in higher solubility of PS. However, the diffusion coefficients in that study differ from those measured by the pulse-field gradient nuclear magnetic resonance (PFG-NMR) methods by more than one order of magnitude.~\cite{rajput2017elucidating}
 The control of clustering may thus represent the  critical objective for battery performance. It is thus of utmost importance to obtain deeper understanding into the structural and transport properties of these highly complex, multi-component PS solutions. 

Our endeavor of this and a preceding study~\cite{park2018molecular} is to develop an efficient yet accurate force field for purely classical molecular dynamics simulations of Li/S battery electrolytes. 
We have successfully described static and transport properties of the solutions of DME/DOL containing LiTFSI and LiNO$_3$ salts.~\cite{park2018molecular}
In this study, we implement PS components into the previously developed model.~\cite{park2018molecular}
For the most relevant PS chain lengths of 4, 6, and 8 sulfur atoms, we investigate clustering behavior of PS, their solvation structure, diffusion, and conductivity properties.

\section{Methods}
\subsection{Model and force fields}
\begin{figure}[ht]
\centering
\includegraphics[width=1\linewidth]{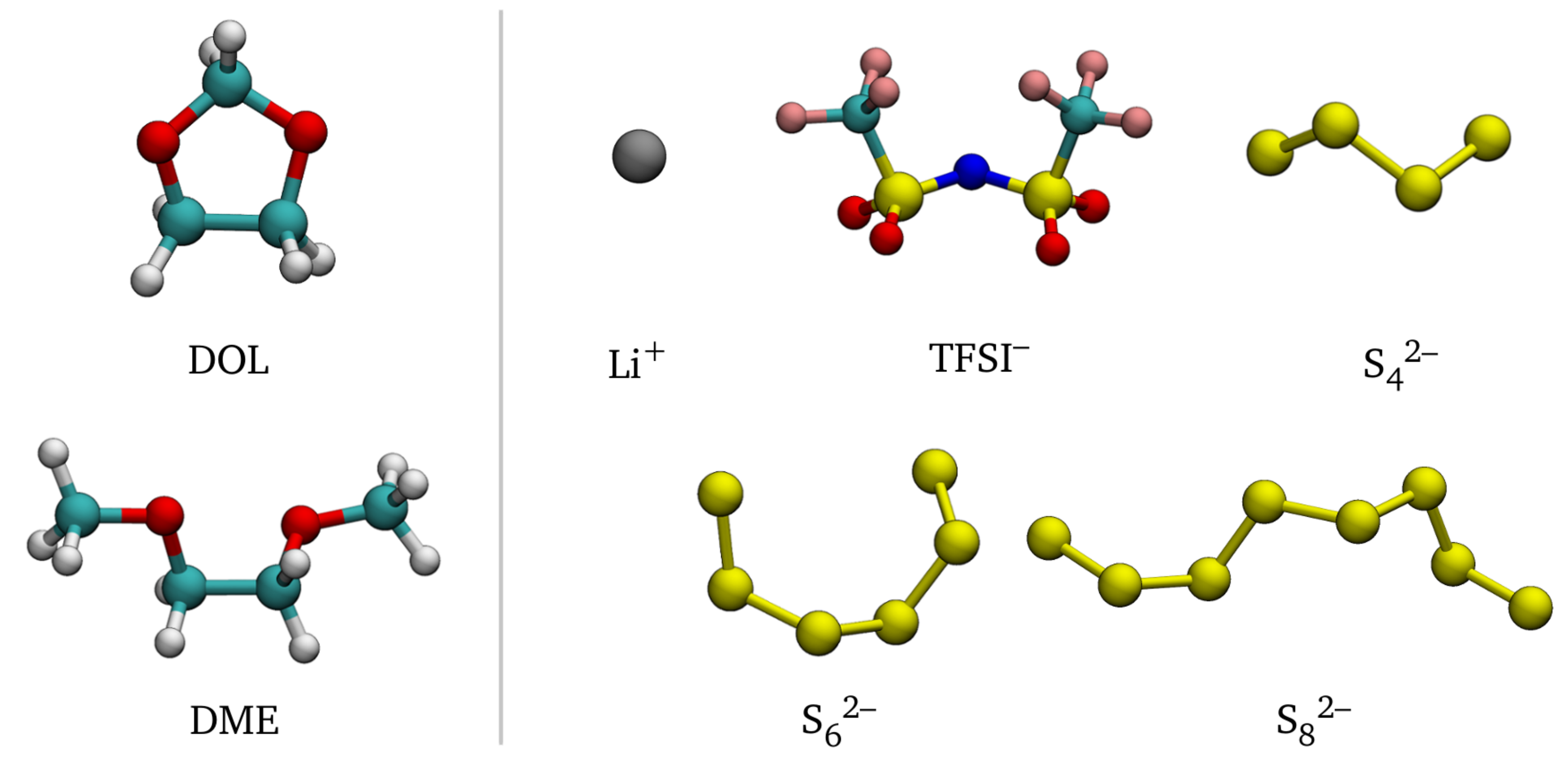}
\caption{Solvent molecules 1,3-Dioxolane (DOL) and 1,2-Dimethoxyethane (DME) and ions Li$^+$,  Bis(trifluoromethanesulfonyl)imide  (\tfsiion), and the polysulfides (PS) with chain lengths $ x $ = 4, 6, and 8 considered in this study.}
\label{fig:snapshots}
\end{figure}

The solvent in our atomistic model is a 1:1 molar mixture of 1,2-Dimethoxyethane (DME) and 1,3-Dioxolane (DOL), which contains different amounts of  \liion and  Bis(trifluoromethanesulfonyl)imide ions (\tfsiion), as well as three different kinds of PS ions:
{\ce{S_{$4$}^{2-}}\xspace}, {\ce{S_{$6$}^{2-}}\xspace}, {\ce{S_{$8$}^{2-}}\xspace} (see \Fig~\ref{fig:snapshots}).

We employ the OPLS-AA force field~\cite{jorgensen1996development, 3anderson} for DME and DOL, the CL \& P force field for \tfsiion,~\cite{ff-tfsi} and Dang \etal~\cite{Dang-Li} for the \liion ion, together with the geometric combination rules for Lennard-Jones (LJ) interactions.
The Coulomb interactions of ions are treated by the electronic continuum correction (ECC) method,~\cite{leontyev2003continuum,leontyev2009electronic,leontyev2010electronicwater,leontyev2010electronic,leontyev2011accounting, leontyev2012polarizable, leontyev2014polarizable} which takes the electronic polarizability into account implicitly. In this approach, formal ionic charges $q_i$ in the interaction Hamiltonian are replaced by effective, rescaled values $q_i^{\mrm{eff}}$,
\begin{equation}\label{eq:qeff}
 q_i^{\mrm{eff}} = \frac{q_i}{\sqrt{\epsilon_{\infty}}}, 
\end{equation}
where $\epsilon_{\infty}=1.93$ is the high-frequency dielectric permittivity of the bare  solvent.~\cite{park2018molecular}
This approach turned out to be very successful for many structural and dynamic properties of LiTFSI/LiNO$_3$ salts in DME/DOL solutions~\cite{park2018molecular} and is applied in this study also to the PS ions.

We devote special attention to the parametrization of the PS ions.
The bonded parameters for PS ions are taken from the recent work by Rajput \etal~\cite{rajput2017elucidating}
In order to obtain the partial charges, we performed quantum mechanical calculations using the electrostatic potential surface method implemented in the GAUSSIAN09 package~\cite{g09} with the B3LYP functional at the aug-cc-pvdz basis set level. The resulting (unscaled) and the corresponding effective (rescaled via \Eq~\ref{eq:qeff}) partial charges of sulfur atoms in PS  are summarized in Table~\ref{table:charges}.

\begin{center}
 \begin{table}[h]
 \caption{Formal ($q_i$) and effective ($q_i^{\mrm{eff}}$, computed via Eq.~\ref{eq:qeff}) partial charges of terminal and internal S atoms in the PS chains. $e$ is the elementary charge.} 
 \centering
  \begin{tabular}
 {@{}| c | c | c | c | c | @{}}
\hline
\hline
    & \multicolumn{2}{c|}{Terminal S} & \multicolumn{2}{c|}{Internal S} \\
    \hline
    & $q_i$ / $e$ & $q_i^{\mrm{eff}}$ / $e$  & $q_i$ / $e$ & $q_i^{\mrm{eff}}$ / $e$ \\
\hline
{\ce{S_{$4$}^{2-}}\xspace} & $-0.7702$ &  $-0.5546$  & $-0.2298$  & $ -0.1655$\\
{\ce{S_{$6$}^{2-}}\xspace} & $-0.6537$ &  $-0.4707$ & $-0.1731$  &  $-0.1247$\\
{\ce{S_{$8$}^{2-}}\xspace} & $-0.6223$ &  $-0.4481$  & $-0.1259$  &  $-0.0907$\\
\hline
\hline
  \end{tabular}
\label{table:charges}
\end{table}
\end{center}

Different LJ parameters for S atoms have been tested using various force fields: OPLS-AA,~\cite{jorgensen1996development} AMBER99,~\cite{sorin2005exploring} CHARMM,~\cite{mackerell1998all} ENCAD,~\cite{levitt1995potential} ECEPP,~\cite{nemethy1983energy} UFF~\cite{rappe1992uff} and DREIDING~\cite{mayo1990dreiding} [see Table~S1 in the Electronic Supplementary Information (ESI)]. 
As it turns out, the LJ  interaction size $\sigma_{\mrm{S-Li}^+}$ between a terminal S atom and Li$^+$ is a critical parameter that determines the degree of aggregation and clustering propensities, as well as conductivity of PS ions.
Therefore, we tuned the LJ parameter manually to $\sigma_{\mrm{S-Li}^+}=  0.275$~nm  in order to reproduce experimental conductivity values. The procedure and the details are provided in ESI (\Figs~S1 and S2).

\subsection{Simulation details and protocols}
All-atom MD simulations are carried out with the GROMACS 5.1 simulation package.~\cite{abraham2016gromacs} 
Initial simulation structures of ions in the solvent are constructed with the PACKMOL package.~\cite{martinez2009packmol} The molecules are randomly inserted into the simulation box and undergo an energy minimization, followed by a simulated annealing from \SI{440}{\kelvin} down to \SI{298}{\kelvin} within a time interval of \SI{3}{\nano\second} in the $NVT$ ensemble. 
The production simulations are carried out in the $NpT$ ensemble at a constant pressure of 1~bar and temperature 298~K. The pressure is controlled by the Parrinello--Rahman barostat with the time constant of 2 ps, whereas the temperature is controlled by the Berendsen thermostat with the time constant 0.1 ps.~\cite{berendsen1984molecular} The LJ potentials are truncated at 1.3 nm. 
The Particle-Mesh-Ewald (PME) method with a Fourier spacing of 0.12 nm and a 1 nm real-space cut-off are used in calculating electrostatic interactions. The LINCS algorithm is used for all bond constraints.

\subsection{Analysis}
\subsubsection{Dielectric constant}
The non-electronic part of the static dielectric constant in the simulations, $\epsilon_{\mathrm{MD}}$, is evaluated from the  fluctuations of the system's dipole moment (excluding ionic monopoles) ${\bf M}_\mrm{MD}$ (using rescaled charges),~\cite{dommert2008comparative, dielectric3}
\begin{equation}\label{eq:dielectric}
   \epsilon_{\mathrm{MD}} = 1 + \frac{1}{3V \epsilon_{0} k_\mathrm{B}T} \bigl( \langle {\bf M}_\trm{MD}^2 \rangle - \langle {\bf M}_\trm{MD}\rangle^2\bigr),
\end{equation}
where $V$, $k_{\mathrm{B}}$, $T$ stand for the volume of the simulation box, the Boltzmann constant, and the absolute temperature, respectively.
Since we are employing the ECC approach, the total effective dielectric constant is 
\begin{equation}
\epsilon=\epsilon_\infty\epsilon_{\mrm{MD}}.
\end{equation}
More details can be found in Ref.~\citen{park2018molecular}.

\subsubsection{Diffusion coefficients}

The self-diffusion coefficients of ions and molecules are computed via the mean square displacement relation,	 
\begin{equation}\label{eq:msd}
 D_\trm{MD} = \lim_{\Delta t \to \infty} \frac{ \big\langle r^2(\Delta t) \big\rangle}{6 \Delta t}.
\end{equation}
In addition, the finite-size effects of the simulation box are corrected by extrapolating measured $D_\trm{MD}$ values at different box sizes $L$ to $L\to\infty$ (see \Fig~S3 in ESI).

 \subsection{Conductivity}
Conductivities of the solutions in our simulations are computed from Ohm's law
\begin{equation}
J = \kappa E, 
\label{ohm}
\end{equation}
where $E$ is an applied electric field and $J=\sum_i J_i  = \sum_i \kappa_i E$ is the resulting total current density. This relation defines partial ionic conductivities $\kappa_i$ for each individual ionic species. The conductivities $\kappa_i$ are evaluated from the linear slope of $J$ versus $E$ in the linear-response regime (see \Fig~S1 in ESI). 

\subsubsection{Clustering}
We define a cluster of PS ions as a group of those PS ions whose at least one of the terminal S atoms is separated from a terminal S atom of any other PS ion in the cluster by less than $r_0=0.53$~nm. The cutoff value $r_0$ is chosen as the first minimum after the main peak of the radial distribution function (RDF) of terminal S atoms and thus corresponds to the distance between two terminal S atoms that have a bridging \liion ion in between (see \Fig~S4 in ESI). 
By ensemble averaging of clusters in the simulations, we obtain an equilibrium cluster size distribution $P(N)$, where $N$ is the number of PS ions in a cluster. The statistical uncertainties are calculated with the block averaging procedure.

\subsubsection{Coordination number and dissociated Li$^+$ ion.}

The coordination number $N_{i,j}$ of the molecule of type $i$ that is surrounded by molecules of type ${j}$ is computed from the RDF as~\cite{HansenMcDonaldbook}
\begin{equation}
 N_{i,j} = 4 \pi c_{j} \int_0^{R_{\trm{M}}} {g}_{{ij}}(r) \, r^2 \, \mathrm{d}r,
\end{equation}
where $R_{\trm{M}}$ is the distance of the first minimum after the first peak in the RDF and $c_{j}$ is the bulk concentration of molecules of the type ${j}$.

\subsection{Experimental Methods}
Electrolyte mixtures were prepared by adding the corresponding stoichiometric ratio of lithium sulfide (\lsone) and elemental sulfur (S$_8$) to a mixture of DME and DOL (1:1 by mole). The uncertainty on the composition of the electrolyte is less than 5\%. Although the various disproportion reactions of lithium polysulfides in solution are known, it is assumed that the majority of PS appears in the length as prepared by the intended stoichiometric ratio, as shown by Barchasz et al.\ using high-performance liquid chromatography (HPLC).~\cite{barchasz2012lithium}
The gravimetric density of the electrolyte mixtures were measured at \SI{18}{\celsius} and 1~bar using a chempro/PAAR DMA 602 density meter. The average value of ten measurements of the natural frequency of a glass tube filled with the solution was taken to calculate the density. Millipore water and air served as reference for this calculation. The error on the values is 2 $\times$ 10$^{-5}$.
The viscosity was determined using a Capillary Viscometer (SI Analytics 50101/0a) and a laboratory stop-watch. The measurement were performed in an argon filled glovebox at \SI{25}{\celsius}.
The error of the measurement varies from 0.07\% to 1\%.
The conductivity was evaluated by performing an impedance spectroscopy in the frequency range of \SI{100}{\milli\hertz} -- \SI{1}{\mega\hertz} with \SI{5}{\milli\volt} RMS voltage signal and 20 points per decade. Three different DC voltages (0, 1, and 1.5~V) were used. A GAMRY interface 1000 potentiostat was used. The cell was a standard CR 2032 coin cell. Two stainless steel plates were used as electrodes. A PTFE-ring (thickness of \SI{0.27}{\milli\meter}, diameter of \SI{10}{\milli\meter}) was used to adjust the distance between the two electrodes. The conductivity was determined by the intersection of the impedance with the $x$-axis at high frequencies in a Nyquist-plot.

\section{Results}
\subsection{Density and dielectric constant}

\begin{figure}[t]
\centering
\subfloat{\label{fig:density-s}\includegraphics[width=1\linewidth]{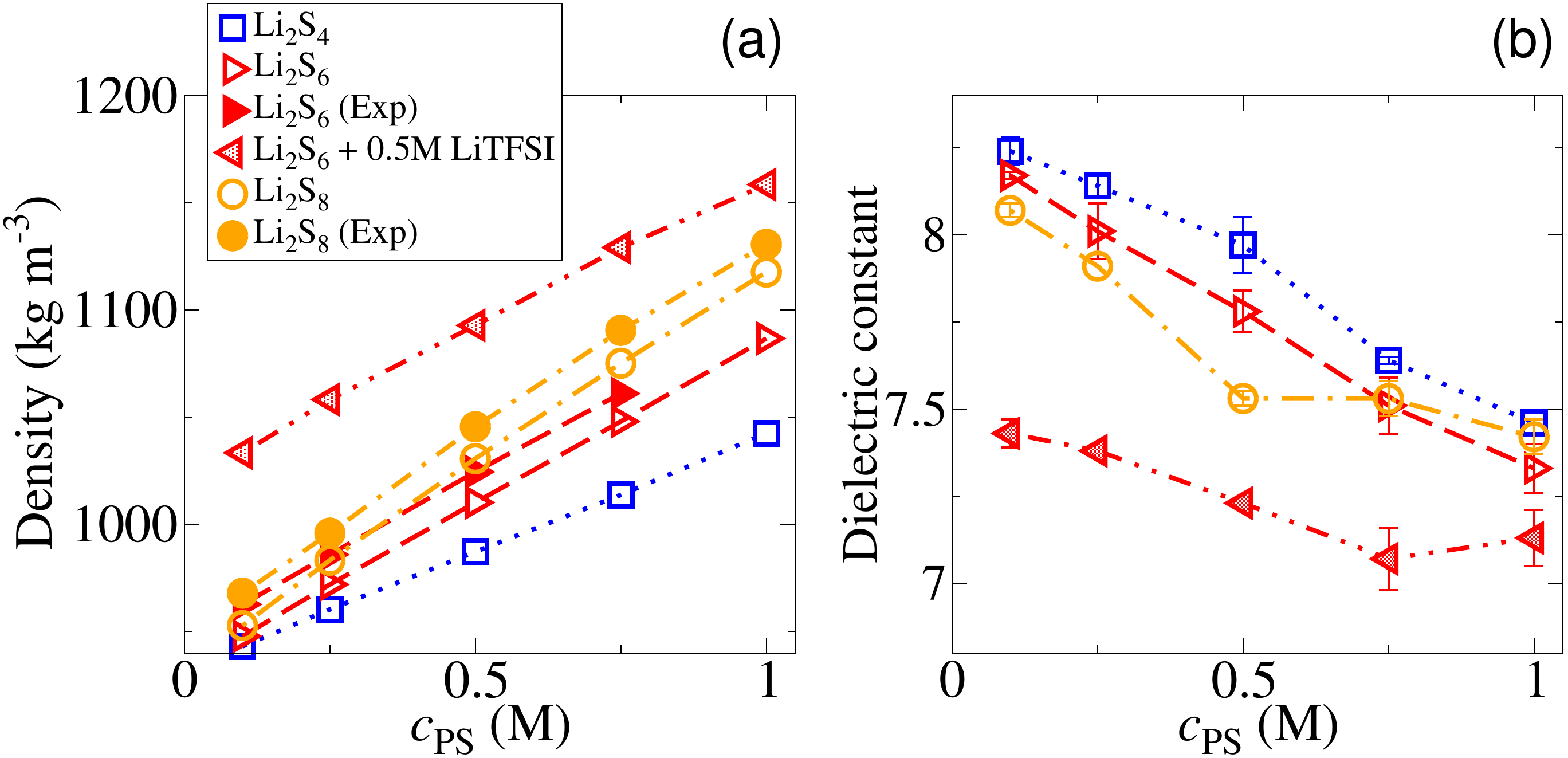}}
\caption{(a)~Density and (b) dielectric constant as a function of the polysulfide concentration in DME/DOL (1:1) solvent in MD simulations. The experimental measurements for Li$_2$S$_6$ and Li$_2$S$_8$ are depicted by solid triangles and circles  in panel (a).}
\label{fig:density-dielectric}
\end{figure}

We start our analysis by calculating the density of various PS in DME/DOL solutions, results of which are plotted  in \Fig~\ref{fig:density-dielectric}a. 
 Universally, the density is an increasing function of the \lsx concentration for all PS types. This rise and the magnitude is in good accordance with our experimental measurements for Li$_2$S$_6$ and Li$_2$S$_8$ (solid triangles and circles, respectively, in \Fig~\ref{fig:density-dielectric}a).
 Note that in our simulations we only consider one-component (i.e., polydisperse) PS solutions. In reality, however, the monodispersity cannot be reached due to the disproportionation reactions of PS, but the majority of the PS should appear in the length as prepared.~\cite{barchasz2012lithium}

The dielectric constant as a function of \lsx concentration is shown in \Fig~\ref{fig:density-dielectric}b. It decreases with ion concentration, as is also the case in other electrolytes.~\cite{ben2011dielectric, glueckauf1964bulk, wei1992ion, chandra2000static} This decrement of the dielectric constant is caused by a local dielectric saturation. Namely, solvent molecules tend to strongly orient and anchor around an ion and do not contribute to the dielectric constant. As more ions are present in the solution, larger fraction of solvent molecules are part of the solvation shells of the ions and thus larger is the local dielectric saturation.
The dielectric decrement is slightly larger for longer PS chains.
The addition of 0.5M LiTFSI into the \lssix solution decreases the dielectric constant considerably, which can be also explained by the local dielectric saturation due to added ions into the solution.

\subsection{Solvation structure and radial distribution functions (RDFs)}\label{section:rdf}
\begin{figure}[ht]
\centering
\includegraphics[width=1\linewidth]{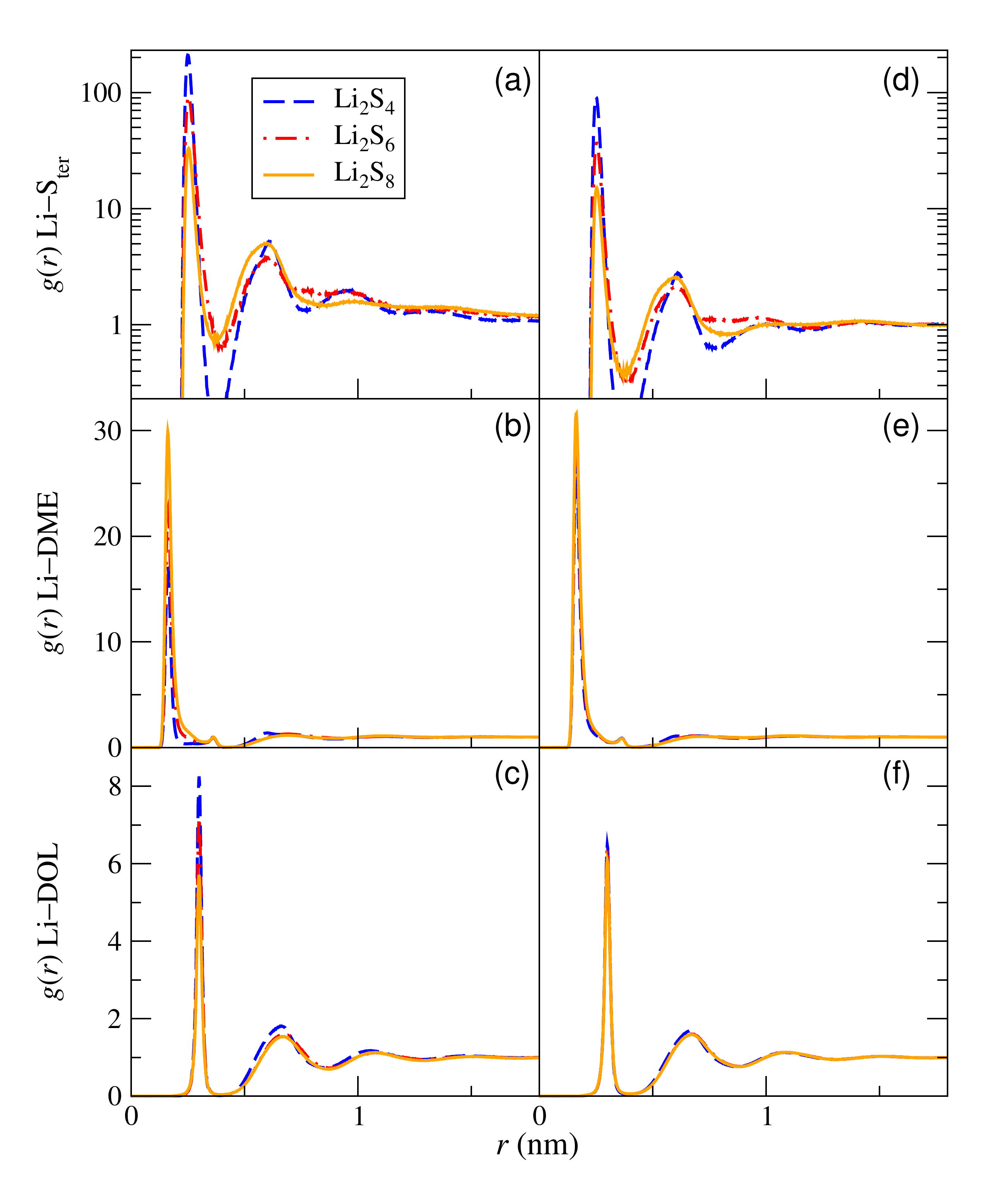}
\caption{Radial distribution functions (at 0.25 M \lsx) between \liion and (a, d)~terminal S (S$_\trm{ter}$) in \sx ($x=$~4, 6, and 8), (b, e)~DME and (c, f)~DOL in DME:DOL (a--c) and in DME:DOL with 1M LiTFSI (d--f).}
\label{fig:output-rdf-li-sul-vs-tfsi}
\end{figure}

\begin{figure}[ht]
\centering
\subfloat[0.25 M \lsfour]{\label{fig:snapshots-s4-2}\includegraphics[width=0.5\linewidth]{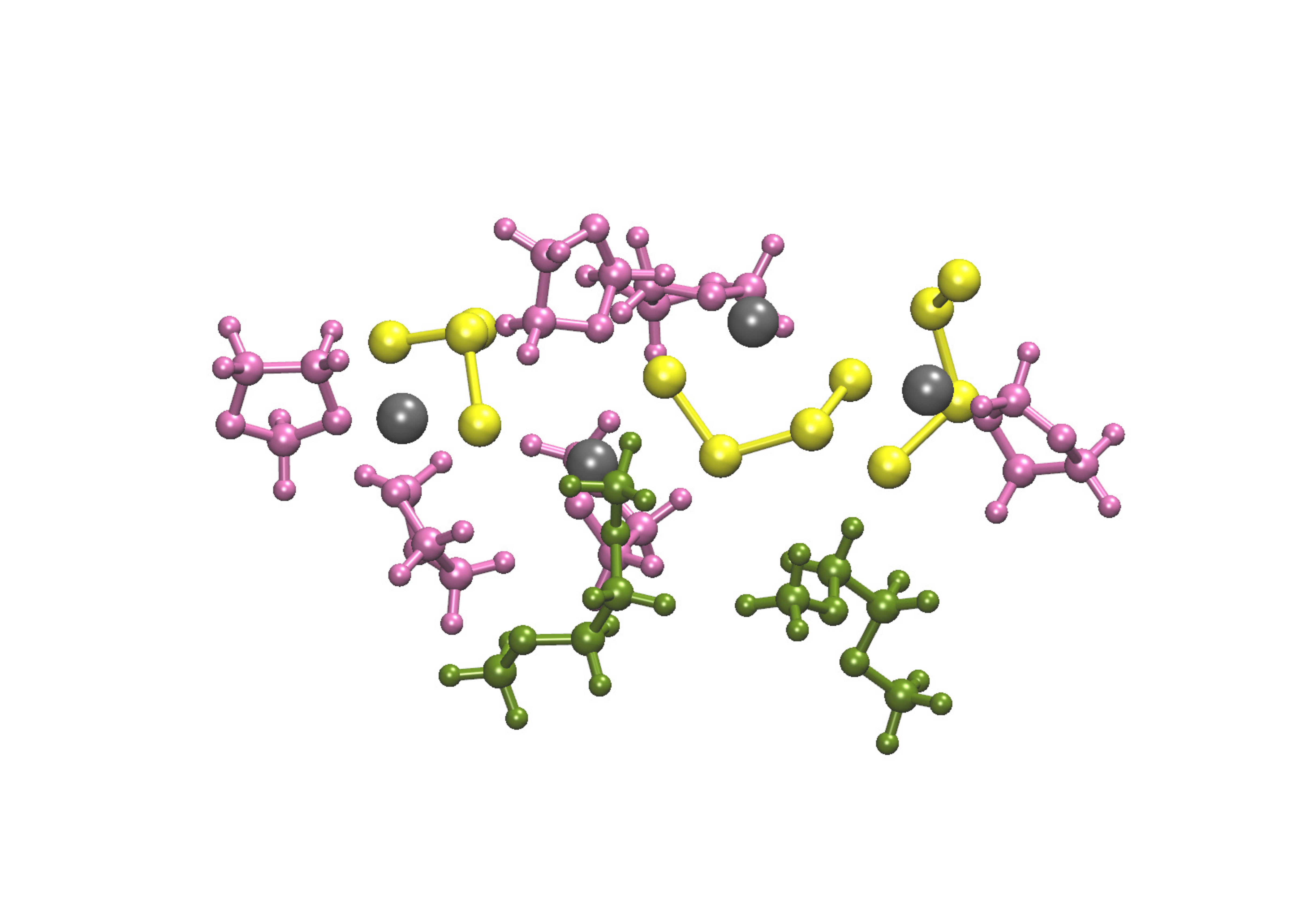}}
\subfloat[0.25 M \lsfour and 1M LiTFSI]{\label{fig:snapshots-s4-2-with-tfsi}\includegraphics[width=0.5\linewidth]{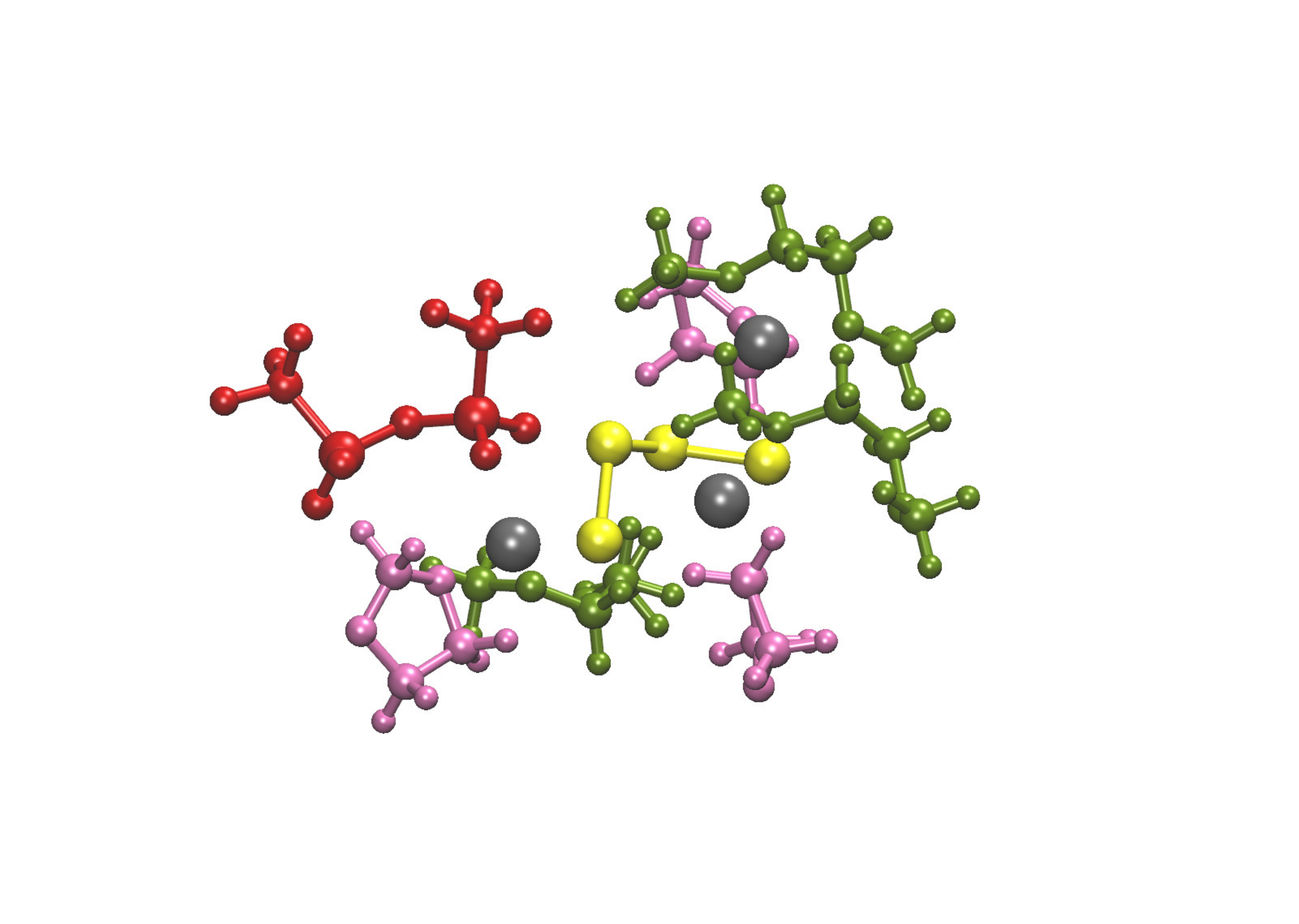}}

\subfloat[0.25 M \lssix]{\label{fig:snapshots-s6-2}\includegraphics[width=0.5\linewidth]{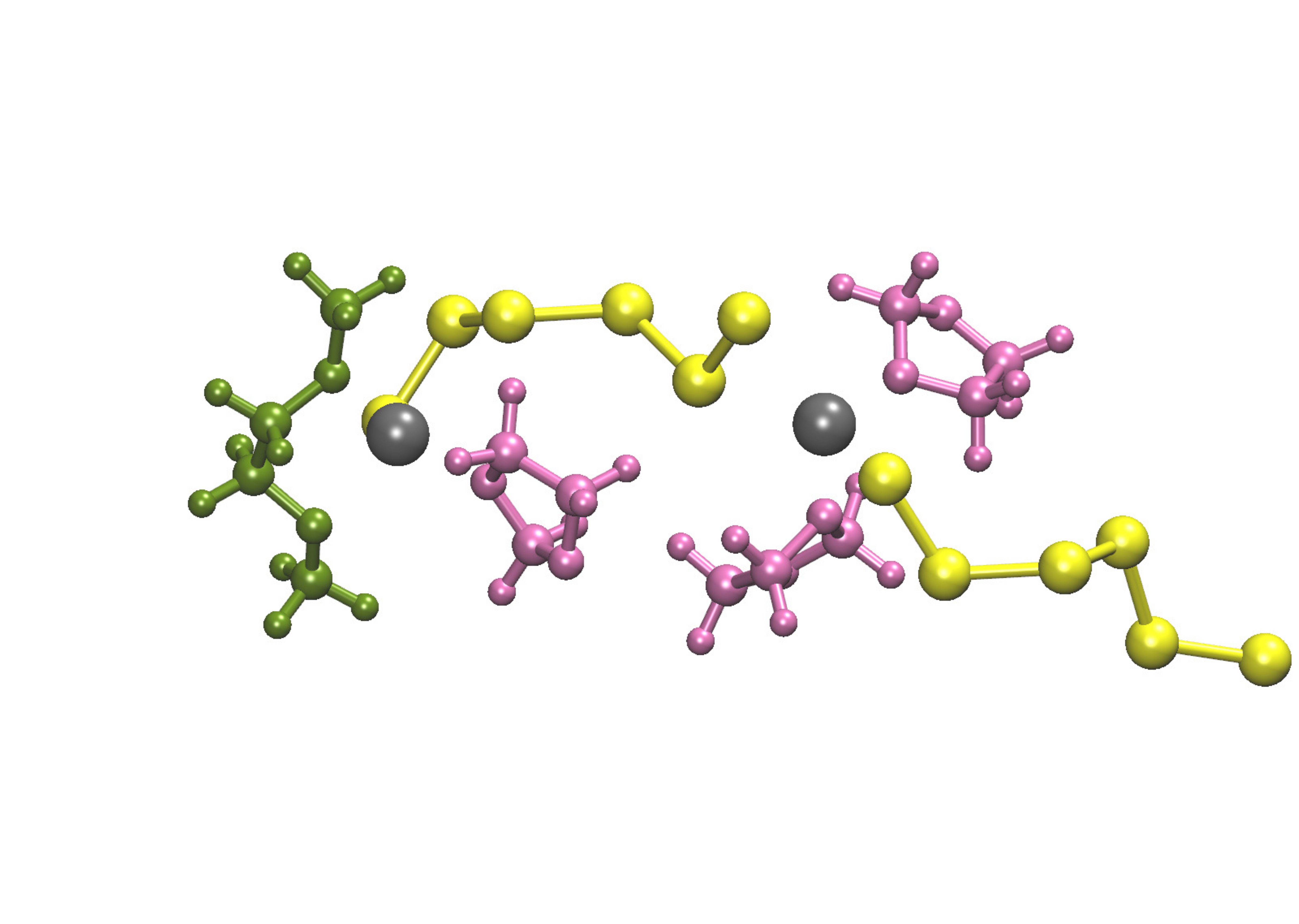}}
\subfloat[0.25 M \lssix and 1M LiTFSI]{\label{fig:snapshots-s6-2-with-tfsi}\includegraphics[width=0.5\linewidth]{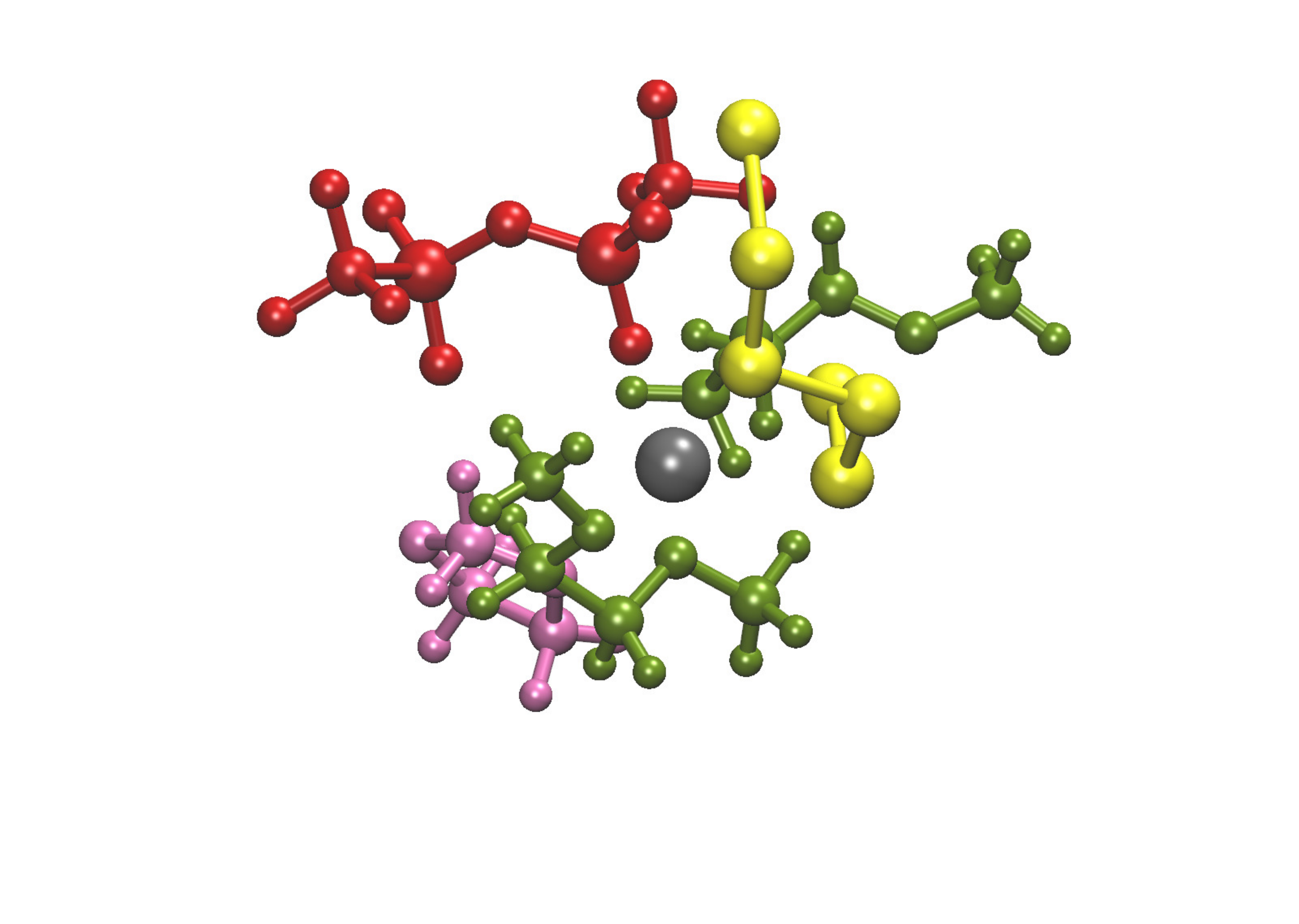}}

\subfloat[0.25 M \lseight]{\label{fig:snapshots-s8-2}\includegraphics[width=0.5\linewidth]{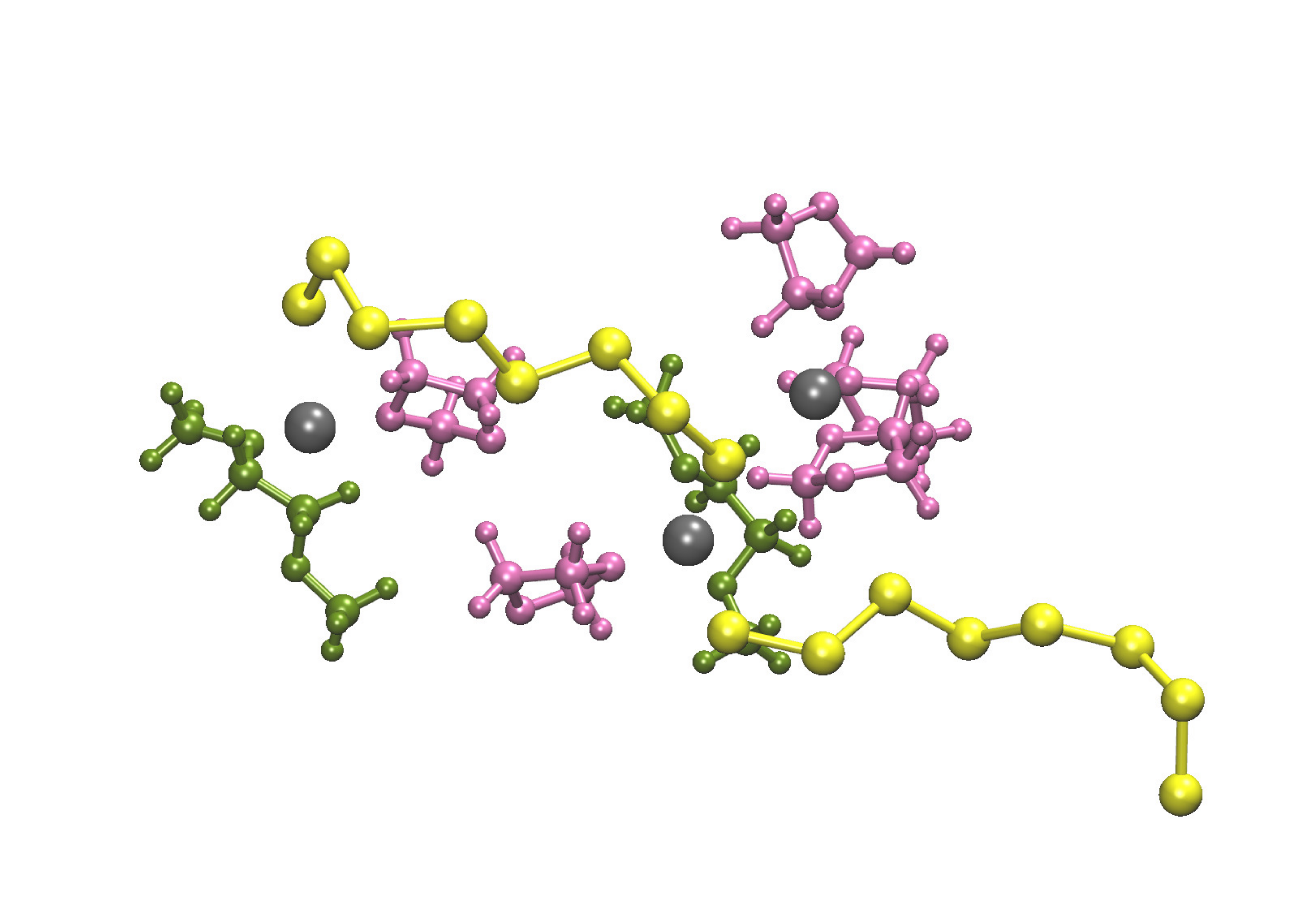}}
\subfloat[0.25 M \lseight and 1M LiTFSI]{\label{fig:snapshots-s8-2-with-tfsi}\includegraphics[width=0.5\linewidth]{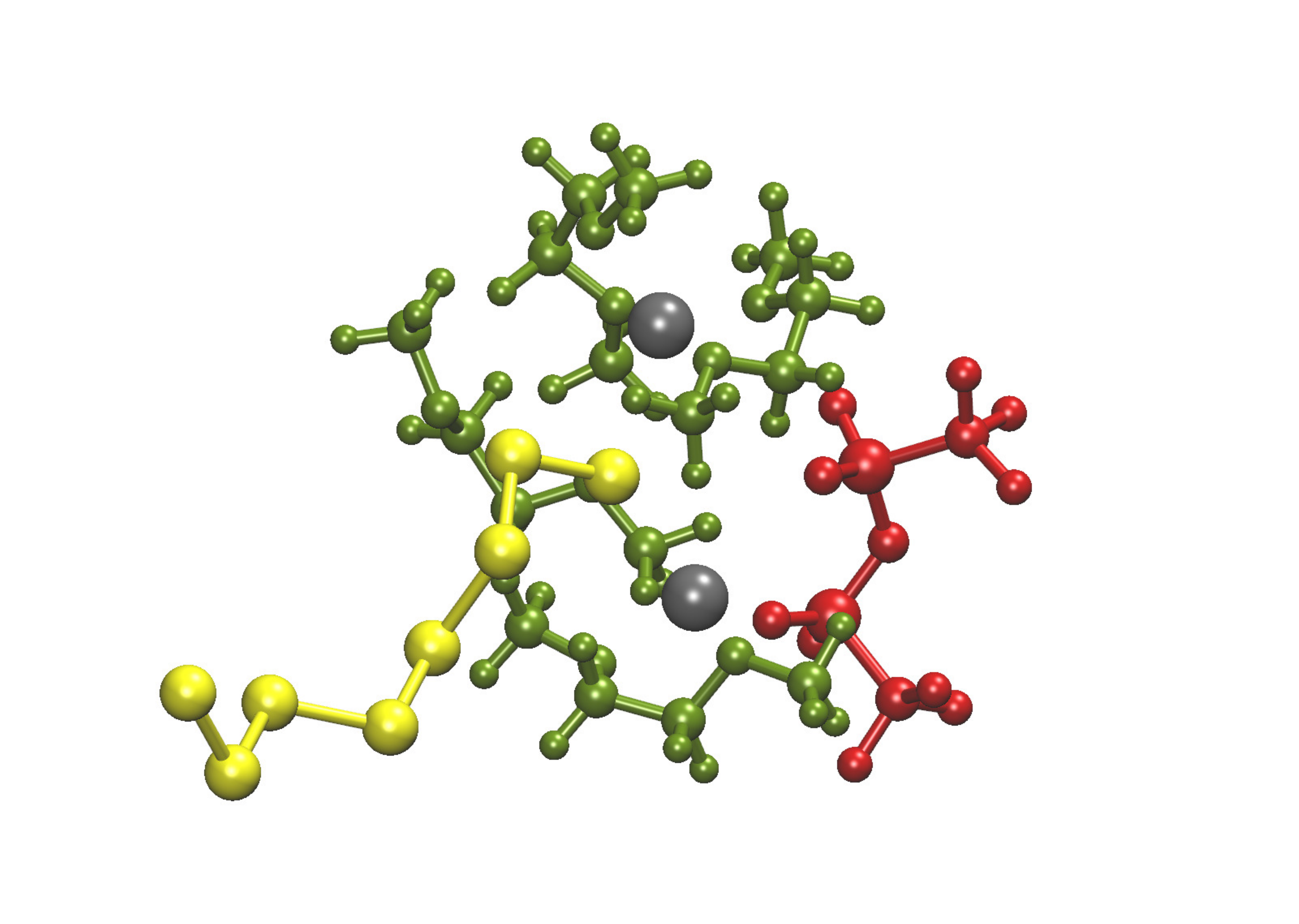}}

\caption{Snapshots of \liion solvation shell in DME:DOL and different amounts of ions. Color code: \liion (gray), \sx (yellow),  \tfsiion (red),  DME (green),  DOL (pink).}
\label{fig:snapshots-solvation}
\end{figure}

In the following, we take a look at the solvation structure of \liion ions. Figure~\ref{fig:output-rdf-li-sul-vs-tfsi} shows RDFs of various molecules and \liion ions in DME:DOL [panels (a--c)] and in 
DME:DOL with 1M LiTFSI  [panels (d--f)].
The top two panels (a, d) show the distribution of terminal S atoms (S$_\trm{ter}$) of PSs.
The very high first peak in all the cases indicates a strong binding affinity between \liion and PS ions and can be attributed to the electrostatic attraction between \liion and \sx. 
Moreover, the height of the peak, and with that the binding strength, are diminishing with the length of the PS ions (when going from \sfour to \seight). These trends are consistent with  a recent classical MD simulation study~\cite{rajput2017elucidating}.
This can be explained by stronger charge localization (of the net valency $-2$) at terminal ends of shorter PS ions.~\cite{kamphaus2017first} As seen in Table~\ref{table:charges}, shorter chains have higher partial charges at the termini, thus facilitating the attraction with \liion.  
Furthermore, the geometry of the PS chains also plays a role in the solvation shell. Snapshots in \Figs~\ref{fig:snapshots-solvation}a, c, e show that short PS chains (e.g., \sfour) are able to tightly wrap around a \liion ion. Conversely, longer PS chains (e.g., \lssix and \lseight) do not pack so tightly around the \liion. 

\begin{figure}[h!]
\centering
\subfloat{\label{fig:coordN-s}\includegraphics[width=1\linewidth]{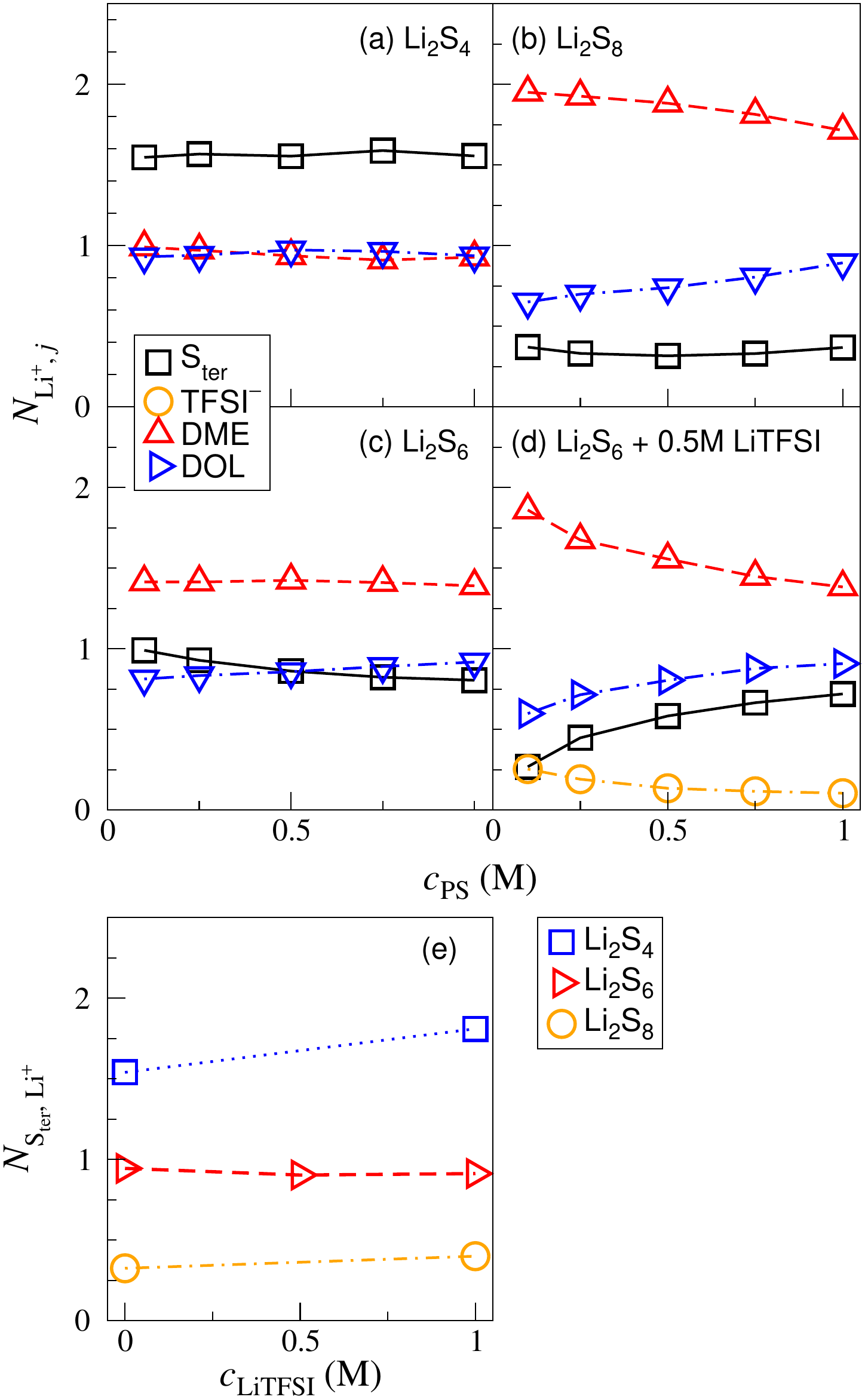}}\\
\caption{(a--d) Coordination numbers of molecules $j$ around the \liion ion ($N_{\ce{Li+}, j}$) as a function of the polysulfide concentration  (see legend). (e)~\liion coordination number around S$_{\mrm{ter}}$  ($N_{\mrm{S}_\mrm{ter}, \ce{Li+}}$) as a function of LiTFSI concentration.}
\label{fig:coordN}
\end{figure}

The different binding strengths result  into different compositions of the \liion solvation shell, which can be described by coordination numbers in \Fig~\ref{fig:coordN}. Namely, shorter PS chains drive out other molecular species from the first hydration shell of \liion [panels (a--d)].
Due to this PS packing in the first solvation shell of \liion, fewer DME or DOL molecules can populate the surrounding of \liion in the presence of \lsfour than in the cases of  \lssix or \lseight (lower peak in  \GDME in the case of \lsfour than in \lseight in \Fig~\ref{fig:output-rdf-li-sul-vs-tfsi}b). The DME coordination number of \liion in Fig.~\ref{fig:coordN} increases from \lsfour to \lseight.   
{\it Ab-initio} MD simulations by Kamphaus~\cite{kamphaus2017first} also showed similar trends. This densely packed solvation structure by \sfour gives less chance for \liion in the solvation shell to contact with solvent molecules. It restricts the \liion exchange between \sfour and solvent molecules, resulting in lower solubility.~\cite{han2017effects, kamphaus2017first}

Now we investigate the effects of LiTFSI in the solution. Figure~\ref{fig:output-rdf-li-sul-vs-tfsi}d shows that the magnitudes of the main peaks in \Glisx decrease after 1 M of LiTFSI is added (cf.\ panel~a), as also consistent with previous studies~\cite{rajput2017elucidating} (also see \GLiTFSI in \Fig~S5 in ESI). 
An important insight can be gained from the \liion coordination number around a PS molecule shown in \Fig~5e. Evidently, the number of \liion ions around S termini does not change upon introducing LiTFSI into the system. This means that LiTFSI does neither weaken the \liion--PS bonds nor do additional \liion ions from LiTFSI bind to PS. Thus the decrease in the RDF peaks (\ref{fig:output-rdf-li-sul-vs-tfsi}d) and the different solvation shell composition (\Fig~\ref{fig:coordN}d) arrive on the expense of the added \liion ions.

\subsection{Conductivity}\label{section:conductivity}
\begin{figure}[h!]
\includegraphics[width=1\linewidth]{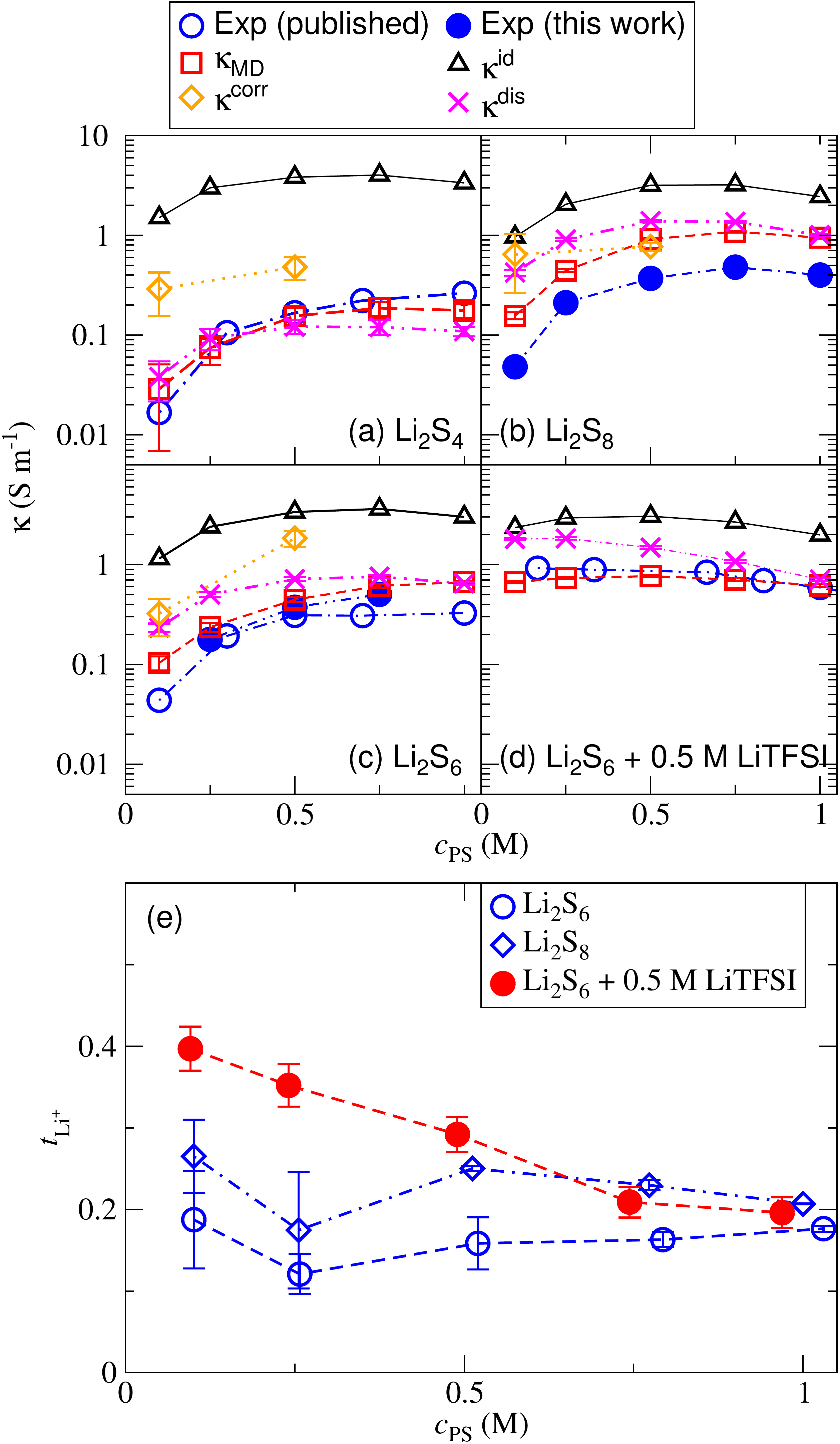}
  \caption{(a--d) Conductivities from MD simulations (squares, triangle left and right), experimental measurements (circles) and the ideal ionic conductivity assuming \Eq~\ref{eq:ne} (triangles up) as a function of the  polysulfide concentration. (a)~\lsfour; exp.\ by Safari \etal,~\cite{Safari} (DME:DOL) (b)~\lseight; exp.\ by us (DME:DOL), (c)~\lssix; exp.\ by Safari \etal~\cite{Safari} (empty circles) and by us (filled circles) (DME:DOL). (d)~\lssix + 0.5 M LiTFSI; exp.\ by Fan \etal~\cite{fan2016solvent} (0.5 M LiTFSI DME:DOL). Diamonds in panels (a), (b) and (c) are $\kappa^{\mrm{corr}}$ using Eq.~\ref{eq:kappa_corr}. Crosses in panels (a), (b), (c) and (d) are $\kappa^{\mrm{dis}}$ using Eq.~\ref{eq:kappa_dis}.
(e) Estimated \liion transference number from MD for different \lsx concentrations (with/without LiTFSI).}
\label{fig:sigma-total}
\end{figure}

In \Fig~\ref{fig:sigma-total} we show the ionic conductivity from the MD simulations and 
 experimental measurements (ours, by Safari \etal,~\cite{Safari} and by Fan \etal~\cite{fan2016solvent}) for PS concentrations in the range of 0.1--1M.
As seen, the experimental trends of the four studied systems ({\ce{LiS_{$4$}}\xspace}, {\ce{LiS_{$6$}}\xspace}, {\ce{LiS_{$8$}}\xspace}, and {\ce{LiS_{$6$}}\xspace}+0.5M {\ce{LiTFSI}\xspace}) are well captured by the simulations. In cases of \lssix\! and \lseight, the conductivities are by a factor of three higher than in experiments. Nevertheless, we regard the results  satisfactory as these quantities are extremely sensitive to the molecular interactions and thus prone to large errors, sometimes of more than an order of magnitude. The conductivities from the  simulations for the ternary electrolytes of \lssix+\,LiTFSI are also congruent with experiments, capturing even the saturation and the decrease in the conductivity with increasing PS concentration above 0.5~M.\\
Ideal ionic conductivity in the limit of low concentrations is given by the Nernst--Einstein (NE) equation
\begin{equation}\label{eq:ne}
 \kappa^{\mrm{id}} = \frac{e^2}{k_{\mrm{B}}T} \sum_i z_i^{2} c_i D_i ,
\end{equation}
where $z_i$ stands for the ion valency and $c_i$ for the ion concentration of species $i$.
Using the diffusivities $D_i$ obtained from the simulations, we calculate the ideal conductivities in \Fig~\ref{fig:sigma-total} (triangles). Clearly, the values are an order of magnitude too high, which we attribute to substantial ionic pairing~\cite{fan2016solvent, park2018molecular}. Namely, \Eq~\ref{eq:ne} is a limiting law and thus neglects ion--ion correlations. 
\\
In a first-order correction to the ideal conductivity, the correlations can be perturbatively taken into account, which leads to~\cite{altenberger1983theory} 
\begin{equation}\label{eq:kappa_corr}
 \kappa^{\mrm{corr}} = \kappa^{\mrm{id}}+ \frac{2e^2}{3 \eta} \sum_{i,j}  z_i  z_j c_i c_j \int_0^{\infty} h_{ij}(r) r \mrm{d}r
\end{equation}
where $\eta$ is the solvent viscosity. 
Here, the ion--ion correlation effects are expressed via $h_{ij}(r)=g_{ij}(r)-1$. Equation~\ref{eq:kappa_corr} thus constitutes a useful structure--transport relationship, applicable at least for not too dense solutions. Taking $h_{ij}(r)$ from our simulations, we plot the correlation-corrected values $\kappa^{\mrm{corr}}$ as diamond symbols in \Fig~\ref{fig:sigma-total}.
Evidently, the negative effect of ionic pairing is qualitatively captured by the correlation term and the values approach closer the MD results. Among all the contributing ion pairs to the second term in \Eq~\ref{eq:kappa_corr}, the most of contribution comes from the \liion--\sx pair (see $\int_0^{\infty} h_{ij}(r) r \mrm{d}r$ values in Table S2). This means that the strong binding between the latter two ions is the main culprit for the observed low conductivity. Still, the correlation correction given by \Eq~\ref{eq:kappa_corr} cannot provide a fair quantitative agreement with the measured values and hence we conclude that even at lower concentrations already correlations beyond the pair level are important, e.g., from clustering effects, see further below. 
\\
Thus, in the limit of very strong ion pairing, where most of the ions are associated into neutral ion pairs and clusters, we can expect that only the dissociated ions contribute to the conductivity. In this simplified picture, we replace ionic concentrations $c_i$ in the NE equation (\ref{eq:ne}) by the concentrations of dissociated ions, $\omega_i c_i$,
\begin{equation}\label{eq:kappa_dis}
 \kappa^{\mrm{dis}} = \frac{e^2}{k_{\mrm{B}}T} \sum_i z_i^{2} \omega_i c_i D_i  ,
\end{equation}
where $\omega_i$ stands for the fraction of dissociated ions of type $i$.
For simplicity of our treatment, we assume that only \liion and PS ions are involved in pairing, whereas  \tfsiion ions are completely dissociated,  $\omega_{\mrm{\ce{TFSI-}}}=1$, as they only weakly bind with \liion.
\liion and PS species are subject to the following equilibrium 
\begin{equation}
\ce{Li2S_{\mrm{\it{x}}}\> \rightleftharpoons \> Li+} + (\ce{LiS_{\mrm{\it{x}}}})^{-}.
\end{equation}
We assume that further dissociation into 2\,\liion + S${_x}^{2-}$ is much less probable and we neglect it.
\begin{figure}[h]
\centering
\subfloat{\includegraphics[width=1\linewidth]{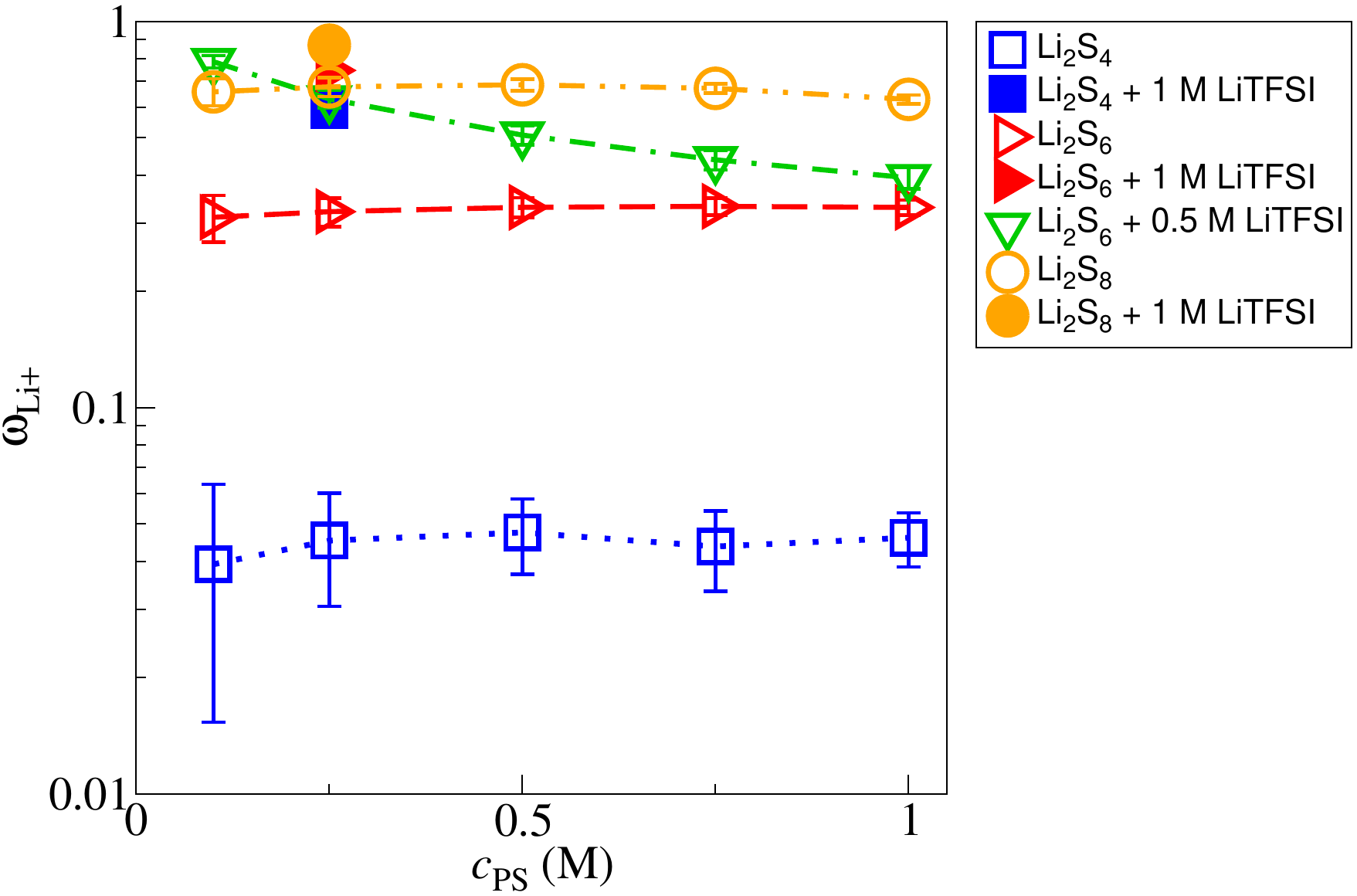}}
\caption{Fraction of dissociated \liion ions as a function of the polysulfide concentration in different solutions. It is defined as the fraction of the \liion population without PS ions in their hydration shells.} 
\label{fig:free-liion}
\end{figure}

Figure~\ref{fig:free-liion} shows the fraction of dissociated \liion in various solutions, defined as the population of those \liion ions that do not have PS in their hydration shells. The dissociation is by far the lowest for \lsfour and higher for  \lssix and \lseight,  similar trends are also observed by a recent study.~\cite{andersen2019structure}
Adding LiTFSI increases the dissociated degree noticeably, which implies that the additional lithium from LiTFSI remains more or less dissociated. 
In cases without LiTFSI, the effective concentrations of \liion and $(\ce{LiS_{\mrm{\it{x}}}})^{-}$
  species are both equal to $2 c_\textrm{PS} \omega_\textrm{\liion}$. 
Introducing LiTFSI adds equivalent concentrations of $c_\trm{\tfsiion}$ to \liion as well as \tfsiion species, since the added \liion are all dissociated. Thus, the modified NE equation then reads
\begin{eqnarray}\label{eq:kappa_dis2}
 \kappa^{\mrm{dis}} = \frac{e^2}{k_{\mrm{B}}T}
&&\! \bigl[(2c_\textrm{PS}\omega_\trm{\liion} +c_\trm{\tfsiion}) D_\trm{\liion} \\
 &&+ 2c_\textrm{PS}\omega_\trm{\liion} D_\textrm{PS}+c_\trm{\tfsiion} D_\trm{\tfsiion}\bigr],\nonumber
\end{eqnarray}
Note that all the species in this treatment are monovalent, therefore $z_i^2=1$.
In \Fig~\ref{fig:sigma-total}a--d, we plot the conductivities  $\kappa^{\mrm{dis}}$ from \Eq~\ref{eq:kappa_dis2} (cross symbols), which in most cases approach much closer to the experimental and MD results than the other two theoretical approaches.
\\
Finally, all three theoretical approaches help us to elucidate the conductivity mechanism of  Li$_2$S$_x$  solutions. Due to high Li--PS pairing, most of the ion carriers are `neutralized' and do not contribute to the conductivity. Only the associated fraction acts on the external electric field, which results into an electric current. When LiTFSI salt is added, it contributes mostly dissociated \liion and \tfsiion ions and therefore fully contribute to the conductivity, which also explains the almost constant trend in \Fig~\ref{fig:sigma-total}d.

As reported by Zheng \etal, \cite{zheng2013obtain} a practical sulfur/electrolyte (S/E) ratio (i.e., density of sulfur in electrolyte)
with improved cycling stability and Coulombic efficiency are achieved for S/E ratio of \SI{50}{\gram \per \liter}. 
This corresponds to approximately 0.4 M \lsfour in our systems for a complete conversion of all sulfur into \lsfour. Above this concentrations, the saturation of the conductivity caused by increasing ionic pairing and viscosity can be one of the limiting factors~\cite{fan2016solvent} for using the high S/E ratio solutions.

Sue \etal~\cite{suo2013new} reported that, in solvent-in-salt systems, high viscosity and incomplete solvation shell facilitate higher \liion transference number ($t_{\trm{\liion}} = J_{\trm{\liion}}/J$), which is unlike in conventional salt-in-solvent electrolytes. In our system, the \liion transference number  does not increase with \lsx concentration (see \Fig~\ref{fig:sigma-total}e).
 Instead, it stays around 0.2 for a wide range of PS concentrations without LiTFSI (as also demonstrated experimentally~\cite{Safari}). The low value can be explained by the fact that \liion ions are mostly moving collectively together with PS ions. The constant value of the transference number is also in accordance with the weak dependence of coordination numbers on PS concentrations. 
 Expectedly, with 0.5M LiTFSI $t_{\trm{\liion}}$ increases, since additional \liion that come from LiTFSI are not bound to PS and contribute to the conductivity to a greater extent. 
 In the latter case, the contribution of PS to conductivity becomes negligible, as evident from the low PS transference number in the presence of LiTFSI (see \Fig~S6a and b). Again, due to the strong binding between \liion and \sfour, \lsfour behaves as a neutral species and is therefore not subject to the electric field. We presume that these short PS chains are more likely to participate in the shuttle mechanism during the charge.

\subsection{Diffusion coefficients}
\begin{figure*}[ht]
\centering
\subfloat{\label{fig:diffusion-1-s}\label{fig:diffusion-s}\includegraphics[width=0.9\linewidth]{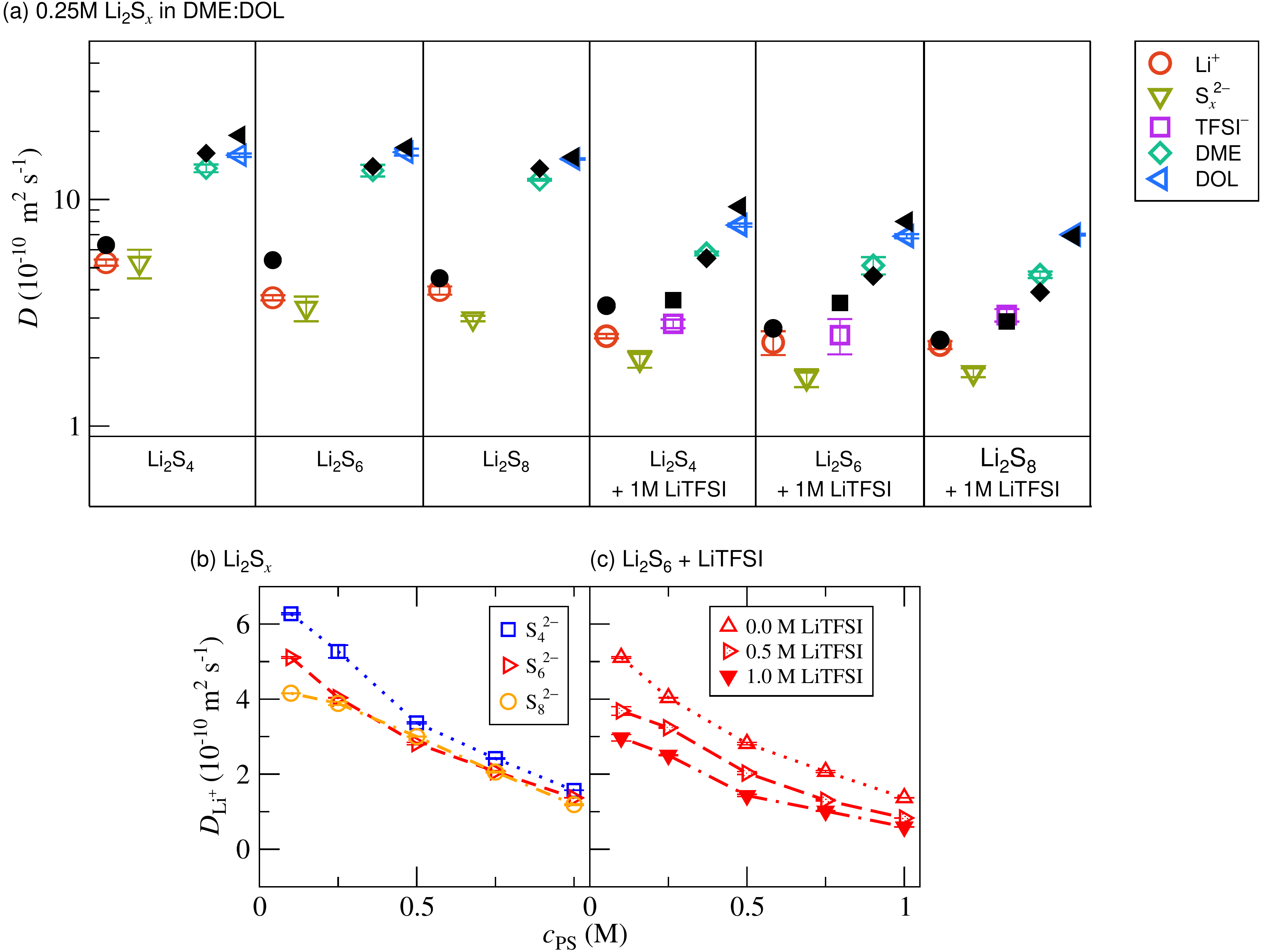}}

\caption{(a) Long-time self-diffusion coefficients $D$ of all species 
in the 0.25 M of the polysulfide as obtained from MD simulations (empty symbols) and PFG-NMR  measurements (black filled symbols).~\cite{rajput2017elucidating} (b)~Diffusion coefficients of \liion in different PS solutions as a function of the PS concentration from MD simulations. (c)~Diffusion coefficients of \liion as a function of \lssix concentration in the presence of different amounts of LiTFSI from MD simulations.} 
\label{fig:diffusion}
\end{figure*}

Evaluating the long-time mean square displacement, we now compute the self diffusion coefficients of all the species in our system. The diffusion coefficients from MD simulations are compared with measured values from PFG-NMR~\cite{rajput2017elucidating} in \Fig~\ref{fig:diffusion}a, which suggests fair agreement.
At this point, we remark that without the ECC treatment of ionic charges, the MD results would  deviate from the experiments by an order of magnitude as we demonstrate in ESI (\Fig~S7). 
Namely, without ECC, the PS--\liion binding is unrealistically strong and thus exaggerates the ion-pair formation.~\cite{pluhařová2013ion, vazdar2013aqueous, pegado2012solvation,mason2012accurate}

Shorter PS chains  (S${_4}^{2-}$) diffuse faster than longer ones (such as S${_6}^{2-}$ and S${_8}^{2-}$), which can simply be explained in terms of the Stokes--Einstein relations, where the diffusion coefficient is inversely proportional to the particle's effective size. 
We furthermore notice (\Fig~\ref{fig:diffusion}a) that the diffusion coefficients of \liion and \sx in the solvent (without \tfsiion ions) are of the same order. This can be related to collective diffusion of \sx and \liion ions due to their strong association. 
As we increase the PS concentration, the \liion diffusivity monotonically decreases (\Fig~\ref{fig:diffusion}b), as also demonstrated experimentally~\cite{Safari}.
This effect can be ascribed to an increasing viscosity in more concentrated PS solutions (\Fig~S8a in ESI). At low concentrations of PS, the diffusion coefficient of \liion significantly depends on the PS type. Namely, shorter PS (\lsfour) promotes higher diffusion than longer PS (\lssix and \lseight). Yet, this differences disappear at higher concentrations of PS (above around 0.5~M).
Interestingly, adding 1 M LiTFSI reduces the diffusion up to about 50\% (\Fig~\ref{fig:diffusion}a, c), as also reported in a previous simulation study~\cite{rajput2017elucidating}.
Also this effect can be attributed to an increased viscosity when LiTFSI is added (see \Fig~S8b in ESI).  
The estimated viscosity of \lssix in DME/DOL with 1M LiTFSI is about a factor of three higher than without LiTFSI. 
As we have seen, viscosity plays a critical role in ionic transport in the PS solutions.  Introducing LiTFSI into dilute PS solutions increases the number of dissociated  \liion and leads to a higher conductivity (\Fig~\ref{fig:sigma-total}). Yet, this ionic effect fades out compared to an increasing viscosity at high PS concentrations.\\

\subsection{Clustering}
\begin{figure*}[ht]
\centering
\subfloat{\label{fig:cluster-log-log-s}\includegraphics[width=1.0\linewidth]{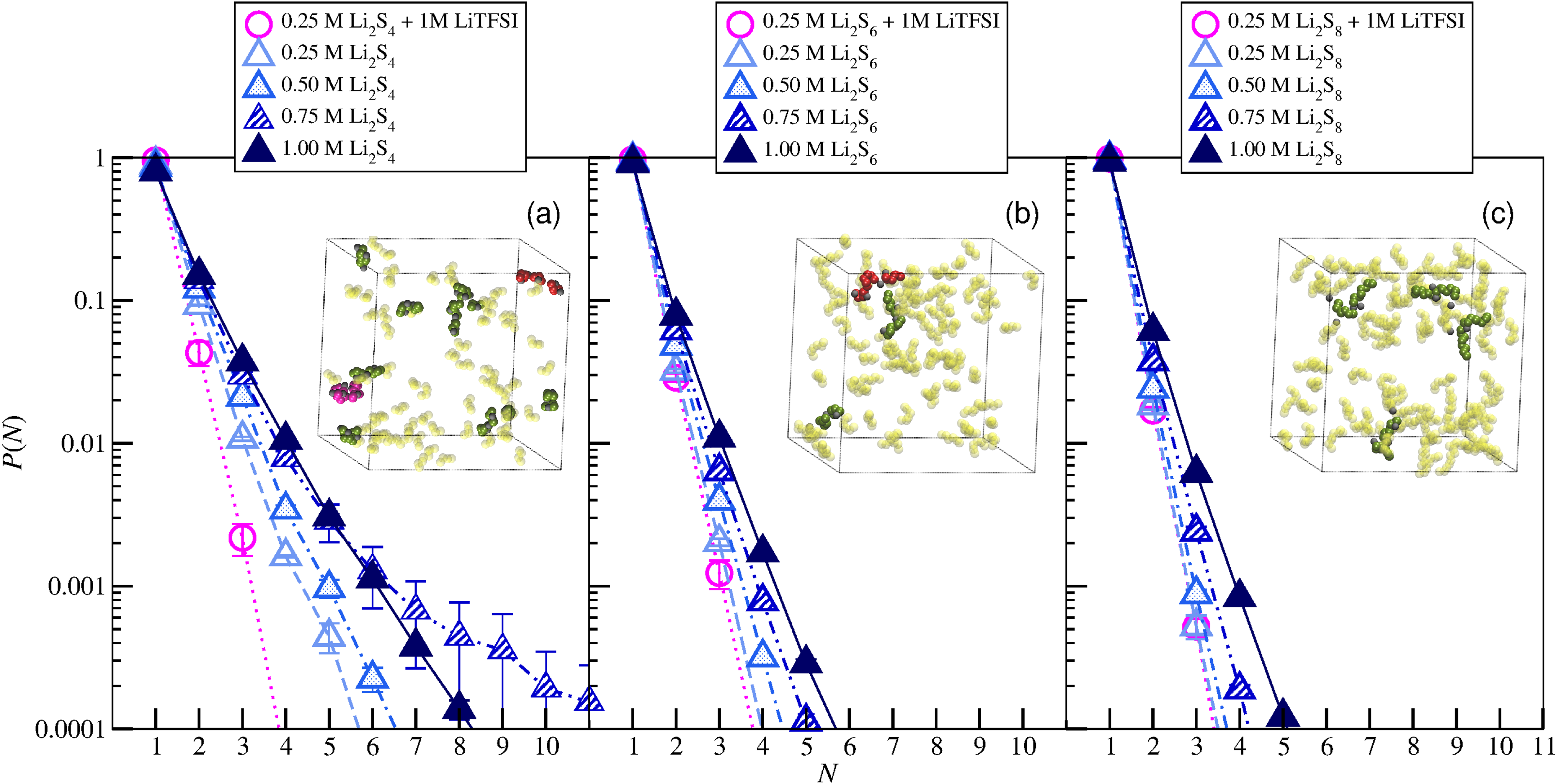}}
\caption{Cluster size distributions for (a) \lsfour, (b) \lssix, and (c) \lseight in DME:DOL.  Insets show representative simulation snapshots of 0.25~M of the polysulfide. Clusters of different sizes are shown in different colors: yellow ($N=1$), green ($N=2$), pink ($N=3$), and red ($N=4$).}
\label{fig:clustering-log-log}
\end{figure*}

Due to the high attraction between terminal sulfur ends of PS and \liion ions, occasionally two different PS ions can bind to the same \liion ion, such that the \liion ion represents a ``bridging'' element for the two PS ions. This can result into supramolecular clusters that are composed of several PS chains and \liion ions.
Figure~\ref{fig:clustering-log-log} shows the cluster-size distribution $P(N)$ of PS in log--lin presentation.
As seen, the clusters do not have a characteristic size but are extremely polydisperse and roughly follow an exponential distribution $P(N)\sim \exp(- N/\overline{N})$, where $\overline N$ is the mean size of the clusters.~\cite{israelachvili2011intermolecular} Some deviations from the exponential behavior occur for larger clusters at higher concentrations, suggesting a cooperative binding. Yet, the cluster size distribution indicates that single monomers and small PS clusters prevail.

Clearly, higher concentrations of PS increase the proportion of larger clusters, simply because the probability of different PS chains to meet is higher in a more concentrated solution.
Even higher concentrations, approaching the solubility limit (not shown), thus provide critical nucleation sites for formation of large cluster precipitates.  The morphology of the cluster may thus depend on the growth and nucleation rates.~\cite{zheng2013obtain}

Moreover, comparing the clustering of different PS lengths reveals that the shorter PSs tend to form larger clusters more readily, especially at high concentrations, compared with longer PSs. A higher frequency of clusters up to 80 atoms was also observed recently for the shorter chains (S$_4^{2-}$) when compared to the longer ones.~\cite{andersen2019structure}
The reason lies in the stronger charge localization at S termini in shorter PS chains, as already discussed above. These results are in line with Vijayakumar \etal~\cite{vijayakumar2014molecular} who reported that \lsfour favors dimer formation, whereas \lseight favors monomer formation in dimethyl sulfoxide (DMSO) solvent, while \lssix being somewhere in between.
Representative simulation snapshots are shown in the insets of \Fig~\ref{fig:clustering-log-log}, featuring larger cluster formation in the case of \lsfour than in the other two cases.
Note that even shorter PS chains, like {\ce{S_{$2$}^{2-}}\xspace}, are even less soluble in the existing solvent and tend to precipitate out of solution even at very low concentrations~\cite{barchasz2012lithium, pascal2017thermodynamic, ding2017situ} (see \Figs~S9 in ESI for the analysis of {\ce{S_{$2$}^{2-}}\xspace} and \Fig~S10 for snapshot). 

Interestingly, adding 1M of LiTFSI inhibits the clustering for \lsfour (\Figs~\ref{fig:clustering-log-log}a--c), as also reported before~\cite{rajput2017elucidating, osellamodelling}. This can, however, not be claimed for the other two PS species.
As discussed above, the presence of LiTFSI does not significantly influence the \liion--PS binding as seen from coordination numbers of S$_\trm{ter}$. However, it apparently tends to inhibit cluster formation via other mechanisms, such as 
increased electrostatic screening due to LiTFSI ions. This influence of the ionic strength on the Li--PS network also impacts the shuttle effect in Li/S batteries.~\cite{kamphaus2017first,peled1989lithium}
Sustaining the Li--S networks by using low ion-pairing salt, can decrease the shuttle effect and increase the cycle performance of the batteries.~\cite{shyamsunder2017inhibiting, chen2016restricting}

\section{Conclusions}

We developed an atomistic model for polysulfides~(PS) in an organic functional solvent that is currently in use for the development of Li/S batteries. 
We focused particularly on structural and dynamic properties of three different sizes of PS ions, {\ce{S_{$4$}^{2-}}\xspace}, {\ce{S_{$6$}^{2-}}\xspace}, {\ce{S_{$8$}^{2-}}\xspace} in the presence of \liion and \tfsiion ions. The conductivity and diffusion coefficients of PS solutions are validated by experimental measurements. Conductivities of PS solutions first exhibit an increase with the {\ce{Li2S_{$x$}}\xspace} concentrations and eventually a saturation at around 0.5~M. The saturation in the conductivity can be linked to viscosity and the ionic correlations between PS and \liion, which also lead to occurrence of supramolecular clusters. 
The tendency of clustering increases with the concentration of {\ce{Li_2S_{$x$}}\xspace} and is more pronounced for shorter PS ions (i.e., $x=4$). Shorter chains have their electronic density more strongly localized at the terminal sulfur atoms than longer chains, thus facilitating the electrostatic attraction with \liion. 
 
The addition of \tfsiion ions leads to larger amount of dissociated \liion ions and to a noticeable increase in the viscosity. The dissociated \liion contributes to the conductivity considerably, on the other hand, increased viscosity inhibits the conductivity at larger PS concentrations. The presence of LiTFSI also reduces cluster formation of shorter PS ions.
 Even though TFSI ions do not significantly reduce the \liion--PS binding, they weaken the binding between multiple PS ions into clusters, partially because of a higher ionic strength. Thus, LiTFSI increases the solubility of PS ions and with that enhances the shuttle effect in the Li/S batteries. 
 
Our simulation results of PS solutions reveal that structural and transport properties are subject to subtle interactions among ions and solvents molecules, which should be considered carefully when it comes to the design of electrolytes for Li/S batteries.~\cite{chaudhari2018assessment} In the next step, studies on different types of anions in Li/PS solutions~\cite{lesch2014combined} as well as in confining electrode materials are envisioned.

\section*{Conflict of interest}
There are no conflicts of interest to declare.


\subsection*{Supporting Information Description}
Force field development for Sulfur non-bonded parameters; Conductivity; Lennard-Jones parameterization based on conductivity; Diffusion coefficients: finite size correction (FSC) (Li$_2$S$_4$ in DME:DOL); Cluster analysis; Radial distribution function (RDF): 0.25 M \lsx in DME:DOL; The integral of the correlation function;  Transference number; Diffusion coefficients without ECC; Viscosity; Cluster analysis and snapshot for \lstwo in DME:DOL. 

\section*{Acknowledgments}
The authors thank Richard Chudoba for useful discussions.
M.K.~acknowledges the financial support from the Slovenian Research Agency (research core funding no.\ P1-0055).


\setlength{\bibsep}{0pt}
\bibliography{literature}


\end{document}